\documentclass[aps,prl,10pt,twocolumn,amsmath,amssymb]{revtex4-1}
\usepackage{graphicx}
\usepackage{array}
\usepackage{xcolor}
\definecolor{green}{rgb}{0.0, 0.5, 0.13}
\DeclareGraphicsExtensions{.png,.eps,.pdf}

\begin{document}

\title{ Topological phase transitions generated by order from quantum disorder }
%{\sl  Substantially revised version of arXiv:1903.11134 }
%\title{ Superfluid roton and new class of
%quantum phase transition tuned by order-from-quantum disorder in a bosonic Quantum Anomalous Hall system }
\author{ Fadi Sun$^{1}$ and Jinwu Ye$^{1,2}$   }
\affiliation{
$^{1}$ The School of Science, Great Bay University, Dongguan, Guangdong, 523000, China \\
%$^{1}$ Institute for Theoretical Sciences, Westlake University, Hangzhou 310024, Zhejiang, China  \\
$^{2}$  Department of Physics and Astronomy, Mississippi State University, MS 39762, USA     }
\date{\today }

\begin{abstract}
    The order from quantum disorder (OFQD) phenomenon was first discovered in quantum spin systems in geometric frustrated lattice.
    Similar phenomenon was also discovered in interacting bosonic systems or quantum spin systems with
    spin-orbit coupling in a bipartite lattice. Here we show that the OFQD also leads to a topological phase transition.
    We demonstrate this new connection in the experimentally realized weakly interacting Quantum Anomalous Hall system
    of spinor bosons in an optical lattice. There are two classes of topological phenomena: the first class is a perturbative one 
    smoothly connected to the non-interacting limit.
    The second one is a non-perturbative one which has no analog in the non-interacting limit.
    Their experimental detections are also discussed.
    %Some conventional OFDQ will not lead to an associated NOFQD phenomenon in the nearby parameter space of a Hamiltonian, but some
    %do lead to an associated NOFQD phenomenon nearby.
    %It is a very general phenomenon and transformative to many other frustrated systems.
\end{abstract}

\maketitle

{\sl 1. Introduction. }---
Searching for new topological phases and topological phase transitions in various materials or artificial systems
are fascinating frontiers in modern condensed matter physics \cite{kane,zhang}.
The quantum anomalous Hall (QAH) is the simplest topological phase with non-vanishing Chern number \cite{QAHthe}.
The non-interacting fermionic QAH has been experimentally realized in Cr doped Bi(Sb)$_2$Te$_3$ thin films
\cite{QAHthe,QAHexp}.  On the other hand, the weakly interacting bosonic analogy of QAH model has also been successfully realized
via spinor bosons $^{87}$Rb \cite{2dsocbec}.
%%and the lifetime of $^{87}$Rb Bose-Einstein condensation have been improved from 300 ms to 900 ms recently.
It is crucial to study the topological properties of bosonic QAH model,
and find possible deep connections between the fermionic QAH model and the bosonic QAH model.

It is well-known that the fermionic QAH model has two bands carrying opposite Chern numbers.
In the non-interacting limit,
when $|h/t|<4$, the corresponding Chern number of the lower and upper band are
$C_{1}=+ \mathrm{sgn}(h)$ and $C_{2}=- \mathrm{sgn}(h)$, respectively;
when $|h/t|>4$,  the corresponding Chern number of the lower and upper bands is $C_1=C_2=0$.
However, the topological properties of weakly interacting bosonic quantum anomalous Hall model
maybe much involved. Due to its bosonic nature, an interaction must be considered at the very beginning.
Unlike the fermionic QAH model,
the weakly interacting  bosonic QAH model is always  in some spin-orbital superfluid phases with the spontaneous U(1) symmetry breaking.
To the quadratic level, there are always Bogliubov quasi-particle excitations above these exotic superfluid phases.
If neglecting the cubic or quartic interactions between them, which is justified near any quantum phase transitions (QPT),
one may ask the question on what are the topology of these Bogliubov quasi-particle bands and possible TPTs.
In this Letter, we map out the Chern number of the bosonic Bogliubov band at the quadratic level.
%is not understand very well,
%because it requires the order from quantum disorder (OFQD) treatment at zero field,
%and nearly order from quantum disorder (NOFQD) treatment at finite field.

In Ref.\cite{NOFQD}, we work out quantum phases and quantum phase transitions in \cite{NOFQD}.
It is the order from quantum disorder (OFQD) phenomena which leads to the quantum ground states at $ h=0 $.
It is the nearly order from quantum disorder (NOFQD) phenomena which leads to the quantum phase transition at $ h=h_c $.
In this work, we focus on the topological aspects of the same system. So the results to be achieved are complementary to those
achieved in \cite{NOFQD}.
We find that there are two kinds of topological phases and TPTs. The first can be considered as the remnants
from the non-interacting fermion limit, so it reduce to this limit smoothly as the interaction gets very small.
It can also be called perturbative regime.
The second has no analog or counterpart in the non-interacting fermion limit,
so it is a completely new feature due to the interaction. We will show that it is completely induced by the
non-perturbative OFQD phenomenon at $ h=0 $ discovered in \cite{NOFQD}. It may also be called perturbative regime.

In addition to the OFQD at $ h=0 $ and the QPT at $ h=h_c $ induced by the NOFQD discovered in \cite{NOFQD},
we find two critical fields $ h_2 > h_1 > h_c >0 $.
At the upper critical field $ h=h_{2} $,
there is a conic band touching at the $ R= (\pi,\pi) $ point of the Brillouin zone (BZ),
so the lower band $ C_1 $ and upper band $ C_2=-C_1 $  Chern number changes from $ (C_1, C_2)=(+1,-1) $ below $ h_{2} $
to  $ (C_1, C_2)=(0,0) $ above $ h_{2} $;
At the lower critical field $ h_{1} < h_{2} $,  there is a conic band touching
at the $ X=(\pi, 0 ) $ and  $ Y=(0, \pi ) $ point of the BZ,
so $ (C_1, C_2)=(-1,+1) $ changes below $ h_{1} $ to  $ (C_1, C_2)=(+1,-1) $ above $ h_{1} $.
Both $ h_1 $ and $ h_2 $ only depends on $ t $ and $ U $, but independent of the SOC $ t_s $,
and approaches to the corresponding non-interacting value $ h_{20}=4t $ and $ h_{10}=0 $ respectively.
We conclude the topology in the regime $ |h| > |h_1 | $ belongs to the first perturbative  class.

We show that $ h=h_{1} > h_c $  which is due to the NOFQD \cite{NOFQD}. When $ h< h_c $, the ground state becomes a XY-CAFM SF phase which breaks
$ [ C_4 \times C_4 ]_D \to 1 $ and leads to  4 bosonic Bogliubov bands in the reduced BZ.
We  find there is always a gap between the second and the third band
and calculate the combined  Chern numbers of the two lower bands and that of the two upper bands  $ (C_{1+2}, C_{3+4})=(-1,+1) $.
So there is no change on the topological band structure across the QPT at $ h= h_c $.
However, the change comes from $ h=0 $ where the OFQD leads to two Dirac points at in-commensurate momenta.
Their positions not only depend on $ t, U $, but also the SOC $ t_s $ and related by the remaining two Mirror symmetries.
As the three parameters change, the two Dirac points approach to each other, then collide at
a degenerate momentum $ \Lambda = (\pi/2,\pi/2 ) $, then bounce off along the normal direction.
Any $ h \neq  0 $ opens the gap on the two Dirac points,
so there is  a TPT from $ h < 0 $ with the combined Chern number  $ (C_{1+2}, C_{3+4})=(1,-1) $ to $ h > 0 $ with  $ (C_{1+2}, C_{3+4})=(-1,1) $,
which is completely induced by the OFQD at $ h=0 $.
We construct effective action to study the TPT and find it always contains an exotic Doppler shift term.
The topology in the regime $ |h| < |h_1 | $ belongs to the second non-perturbative  class.
We also critically comment on the common conceptual mistakes made in the previous theoretical or experimental literatures
to attempt to evaluate edge modes within the bulk gaps associated with the bosonic bulk Chern numbers.
We also elucidate the physical meanings of the combined Chern numbers.
Finally, we discuss the experimental detection of these topological phenomena, especially the ones induced by the OFQD near $ h=0 $.

{\sl 2. The Hamiltonian, quantum phases and quantum phase transitions (QPT). }---
The recently experimentally realized two-component spinor Bose-Hubbard Hamiltonian with  a spin-orbit coupling
\cite{2dsocbec} can be written as
\begin{align}
	&\mathcal{H} =
	-\!\sum_i[a_i^\dagger (t\sigma_z\!-\!it_\text{s}\sigma_x) a_{i+x}
		 \!+\!a_i^\dagger (t\sigma_z\!-\!it_\text{s}\sigma_y) a_{i+y}\!+\!h.c.]
		\nonumber   \\
	&\qquad
	-h\sum_i(n_{i\uparrow}-n_{i\downarrow})
	+U\sum_i n^2_i -\mu\sum_i n_i\>.
\label{QAH0}
\end{align}
where $\sigma=\uparrow$ or $\downarrow$ denotes the $^{87}$Rb atoms
in the state $|1,m_F=0\rangle$ or $|1,m_F=-1\rangle$, respectively.
Since the experiment can only achieve relatively weak spin-orbit coupling,
our discussion will focus on regime $|t_s|<2|t|$.

The quantum phase diagram of Eq.\eqref{QAH0} was studied in Ref.\cite{NOFQD}.
Especially, we found a new phenomenon we named Nearly order from quantum disorder (NOFQD)
which captures the  delicate competition between the effective potential generated by  the order from quantum disorder (OFQD)
and the Zeeman field. It is this competition which splits a putative first order quantum phase transition (QPT) at $ h=0 $ into two second order
QPTs at $ h = \pm h_c $.
Here we briefly summarize the main results:
when $ h > h_c $, it is in the Z-FM superfluid
with the condensate wavefunction $\Psi\propto\chi_{\uparrow}$;
when $ h < h_c $, it is the  XY-CAFM superfluid
with the condensate wavefunction
$\Psi\propto\cos(\theta/2)\chi_{\uparrow}
+e^{i(\mathbf{Q}\cdot\mathbf{r}+\phi)}\sin(\theta/2)\chi_{\downarrow}$,
where $\mathbf{Q}=(\pi,\pi)$ is the orbital ordering wave-vector, $\theta=\arccos(h/h_c)$, and $\phi=\pm\pi/4,\pm3\pi/4$.
The $h_c$  was estimated as $h_c\sim n_0U^2t_s^2/t^3$.
%In the following, we will calculate the band Chern number for the bosonic quantum anomalous Hall model.

However, Ref.\cite{NOFQD} fucus only on the quantum phases and QPTs,
ignored the topological features of the Bogoliubov bands, also did not touch
what are the connections of these quantum phases and QPTs to the topological nature of the QAH Hamiltonian.
This work will address these outstanding open problems.
Without loss of generality, we can set $h,t,t_\text{s}>0$
and then discuss the topology of the Bogoliubov excitation bands at $h>h_c$ and $h<h_c$ separately.

{\sl 3. Evaluating the Chern number of the bosonic quasi-particle band from the lattice theory. }---
To exam the topology of the Bogoliubov excitation bands,
we need not only the eigenvalues but also the eigenvectors,
thus one need also to pay attention to the Bogoliubov transformations. %used in the main text.
After replacing the bosonic operator by its average plus a quantum  fluctuation,
we obtain a quadratic Bogoliubov Hamiltonian via expanding the Hamiltonian to the second order in the quantum fluctuations.
Due to the spontaneous U(1) symmetry breaking,
the quadratic Hamiltonian in $\mathbf{k}$-space, $H_\mathbf{k}$, is a $2N\times 2N$ matrix,
\begin{align}
    \mathcal{H}=E_0+\frac{1}{2}
    \sum_\mathbf{k} (\psi_\mathbf{k}^\dagger,\psi_{-\mathbf{k}})
    H_\mathbf{k}
    \begin{pmatrix}
        \psi_\mathbf{k}\\
        \psi_{-\mathbf{k}}^\dagger\\
    \end{pmatrix}\>,
\label{eq:Hb}
\end{align}
where $\psi_\mathbf{k}^\dagger\!=\!(\psi_{1,\mathbf{k}}^\dagger,\cdots,\psi_{N,\mathbf{k}}^\dagger)$.
The $1,\cdots,N$ indices count the spin degrees of freedom
and also the momentum resulting from the spontaneous translational symmetry breaking i9n the ground state.
As shown in \cite{NOFQD}, when  $h>h_c$, the ground state is the Z-FM phase,
%Eq.M(3) gives
we have $N=2$ and $\psi_\mathbf{k}^\dagger
=(\psi_{\mathbf{k}\uparrow}^\dagger,\psi_{\mathbf{k}\downarrow}^\dagger)$;
when $h<h_c$, the ground state is the XY-AFM phase,
%Eq.M(9) gives
we have $N=4$ and  $\psi_\mathbf{k}^\dagger=
(\psi_{\mathbf{k}\uparrow}^\dagger,\psi_{\mathbf{k}\downarrow}^\dagger,
\psi_{\mathbf{k}+\mathbf{Q}\uparrow}^\dagger,\psi_{\mathbf{k}+\mathbf{Q}\downarrow}^\dagger)$. where $ \vec{Q}= ( \pi, \pi) $.

Diagonalizing Eq.\eqref{eq:Hb}  by a $2N\times2N$ Bogoliubov transformation matrix $T_\mathbf{k}$, so that
$T_\mathbf{k}^\dagger H_\mathbf{k} T_\mathbf{k}
=\mathrm{diag}(\omega_{1,\mathbf{k}},\cdots,\omega_{N,\mathbf{k}},
\omega_{1,-\mathbf{k}},\cdots,\omega_{N,-\mathbf{k}})$ where $\mathrm{diag}(\cdots)$ means a diagonal matrix with diagonal elements $\cdots$,
and $ (\begin{smallmatrix}
        \psi_\mathbf{k}\\
        \psi_{-\mathbf{k}}^\dagger\\
    \end{smallmatrix})
    =T_\mathbf{k}
    (\begin{smallmatrix}
        \alpha_\mathbf{k}\\
        \alpha_{-\mathbf{k}}^\dagger\\
    \end{smallmatrix})  $, we obtain
\begin{align}
    \mathcal{H}=E_0+\sum_{n,\mathbf{k}}\omega_{n,\mathbf{k}}(\alpha_{n,\mathbf{k}}^\dagger \alpha_{n,\mathbf{k}}+1/2),
\end{align}
where $ \mathbf{k} $ sums over the corresponding Brillouin Zone (BZ),
$\omega_{n,\pm \mathbf{k}} > 0 $, $n=1,\cdots,N$ are the $N$ Bogoliubov excitation bands.

In contrast to the fermionic cases, to keep the bosonic commutation relations,
the $T_\mathbf{k}$ is required to be a para-unitary matrix instead of a unitary one, which means
$T_\mathbf{k}^\dagger \tau_3 T_\mathbf{k}
=T_\mathbf{k} \tau_3 T_\mathbf{k}^\dagger=\tau_3$
and $\tau_3=\sigma_z\otimes I_{N\times N}$.
The Berry curvature of the $n$-th Bogoliubov excitation band can be calculated via
$F_n(\mathbf{k})=i[\epsilon_{\mu\nu}\tau_3 \partial_\mu T_\mathbf{k}^\dagger \tau_3 \partial_\nu T_\mathbf{k}]_{nn}$,
where $\mu,\nu=k_x,k_y$.
The Chern number of the $n$-th Bogoliubov excitation band can be evaluated via a integral
\begin{align}
    C_n=\frac{1}{2\pi}\int_\text{BZ} d^2 \mathbf{k} \> F_n(\mathbf{k})\>,
\label{eq:C-n}
\end{align}
where the integral is over the corresponding BZ.

{\sl 4. Evaluating the Chern number of the bosonic quasi-particle band from the continuum theory. }---
Since the band closing is the signature of a topological phase transition (TPT),
the change of band Chern number may also be computed in terms of a continuum theory.

Near the band touching point, the typical effective Hamiltonian can be expressed in terms of the polar coordinate $(k,\xi)$ with $k_x=k\cos\xi$ and $k_y=k\sin\xi$,
\begin{align}
    H_\mathbf{k}=v_x k^n\cos(n\xi) \sigma_x+ v_y k^n\sin(n\xi) \sigma_y- \Delta\sigma_z
\label{Hn}
\end{align}
where $|v_x|=|v_y|=v$ and $n=1,2,3,\cdots$ stands for the order  ( or charge ) of the touching point.

The corresponding dispersion relation is
$\epsilon(\mathbf{k})=\pm\sqrt{\Delta^2+ v^2 k^{2n}}$,
and $\epsilon(\mathbf{k})\sim k^n$ as $\Delta\to0$ approaching the TPT.
The Berry curvature of the lower branch is
$F_-(\mathbf{k})=-v_xv_y\Delta  k^{-2+2n} n^2/|\epsilon(k)|^3$,
thus its Chern number is
\begin{align}
    C_-=\frac{1}{2\pi}\int d^2\mathbf{k} F_-(\mathbf{k})
     =-\frac{n}{2}\mathrm{sgn}(v_xv_y\Delta)\>.
\label{C-}
\end{align}
where the $ sgn $ indicates the chirality of the Dirac boson.
Therefore the band touching through the changing sign of $\Delta$ from positive to negative across the TPT
leads to a change of Chern number $ \Delta C_=-n\, \mathrm{sgn}(v_xv_y) $.
%\label{deltaC-}
%\end{align}
When there are multiple band touching points,
the total change of Chern number needs to sum up all
the contributions from all the band touching points.

\begin{figure}[!htb]
    \includegraphics[width=\linewidth]{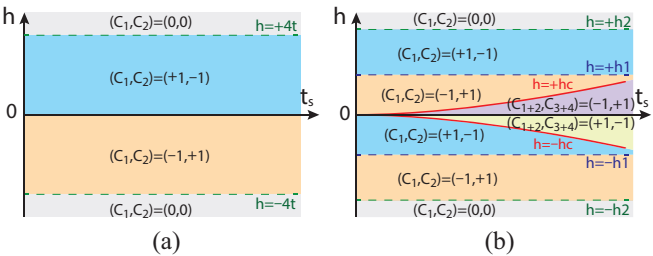}
    \caption{ The band Chern numbers of (a) The non-interacting fermionic QAH,
    and (b) The weak-interacting bosonic QAH.
    The topological non-trivial regions ($C_i\neq 0$) in (b) is slightly larger than (a).
    The regime $ | h | >  h_1 $ region is smoothly connected to the non-interacting limit in (a), so can be called perturbative region.
    While the entire $ | h | < h_1 $ region is induced by the non-perturabative OFQD phenomenon at $ h=0 $.
    It has no non-interacting analog, so may be called the  non-perturabative region which shrinks to zero as $U$ goes $0$.  }
\label{CN}
\end{figure}

In the superfluid phases, due to the BEC at $ k=0 $,
the Bogoliubov transformation matrix $T_k$ diverges at $k=0$,
the Chern number of the lowest band $C_1$ may not be well-defined,
but those of the higher bands $C_{n>1}$ are usually well-defined, and quantized to be integers.
\cite{Ueda2015}.
In this Letter, we are still able to calculate $C_1$ by excluding
the momentum $k$ where the BEC resides.
The results are summarized in Fig. \ref{CN}.
When comparing  the non-interacting fermionic QAH model with the interacting bosonic QAH model,
we find that the bosonic interaction leads to new and additional patterns of topological bands.
In the following sections, we present the details of calculations leading to Fig.\ref{CN}.

\begin{figure}[!htb]
    \includegraphics[width=\linewidth]{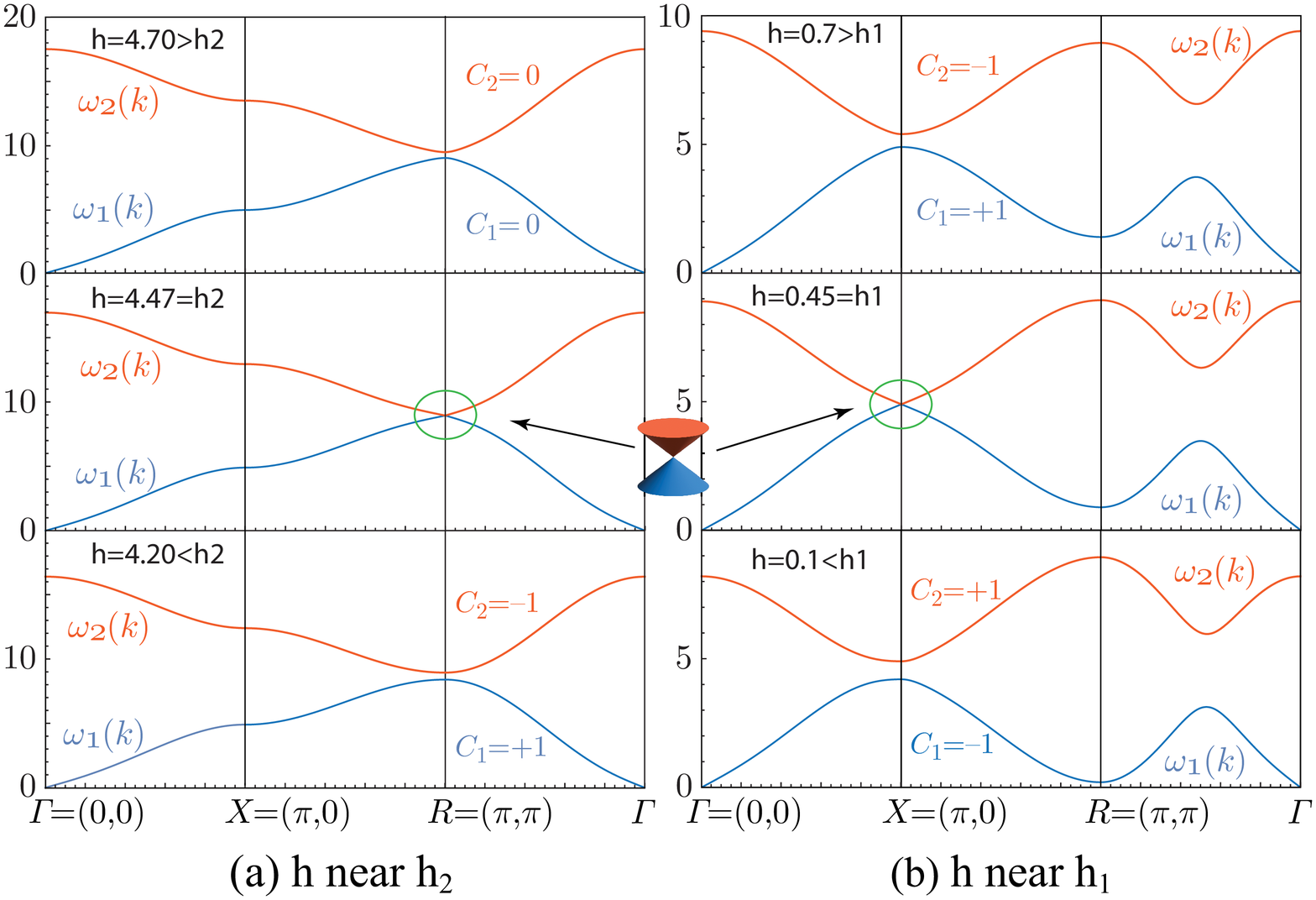}
    \caption{The perturbative topological band structure of Bogoliubov excitation bands at $h>h_c$
    and near (a) $h_2$ and (b) $h_1$. It is smoothly connected to the non-interacting limit.
    The parameters are $t=1$, $t_s=1/2$, $n_0U=1$, thus $h_2\approx4.47$ and $h_1\approx0.48$.
    When $h=h_2$,
    there is a Dirac conical band touch at the R point $k_R=(\pi,\pi)$ of the Brillouin zone (BZ).
    When $h=h_1$, there is a Dirac conical band touch at the X point  $k_X=(\pi,0)$ of the BZ.
    The $[C_4\times C_4]_\text{diag}$ symmetry
    tells there is also a Dirac conical band touch at the Y point $k_Y=(0,\pi)$ of the BZ.  }
\label{Band-h}
\end{figure}

{\sl 5. The topology of the Bogoliubov band when $ h > h_c $.}---
In the $h>h_c$ case, the ground state is the Z-FM phase,
%this corresponds to Eq.(3) and Eq.(4) in Ref.\cite{NOFQD}.
there are  $N=2$ energy bands,
the Bogoliubov transformation matrix $T_k$ is a $4\times4$ matrix.
The corresponding band structure %in Eq.M(4)
is plotted in Fig.\ref{Band-h} where
\begin{align}
   h_{2}    & =\sqrt{4t(4t+n_0U)}    \nonumber  \\
   h_{1}  & =\sqrt{2t(2t+n_0U)}-2t
\label{h12}
\end{align}

When $h=h_{2} $, the two bands conically touch at $k_R=(\pi,\pi)$;
when $h=h_{1} $, they conically touch  at $k_X=(\pi,0)$ and $k_Y=(0,\pi)$.
%Each conical band touch suggests the change of band Chern number is 1,
%thus the change of band Chern number across $h_{1}$ and $h_{2}$ are $2$ and $1$ respectively.
A direct evaluation of Eq.\eqref{eq:C-n} on the lattice shows that:
when $ h > h_{2} $, the lower band and the upper band Chern number are $C_1=C_2=0$ respectively;
when $h_{1}<h<h_{2}$,  $ ( C_1, C_2)=(1,-1) $;
when $h<h_{1}$,  $ ( C_1, C_2)=(-1,1) $.
%Eq.\eqref{C-} tells each conical band touch leads the change of Chern number  $+1$ or $-1$
%depending on its chirality.
%So the numerical results are consistent with our prediction of the change of band Chern number.
Below, we construct continuum theories to confirm the change of the band Chern number.

At $h=h_2$, one needs to expand the Hamiltonian around $k_R=(\pi,\pi)$,
the $T_k$ at $k_R$ tells the eigenmodes are
$\alpha_{1,R}=
(1/\sqrt{8})[2(4+n_0U/t)^{-1/4}+(4+n_0U/t)^{1/4}]\psi_{R\uparrow}
-(1/\sqrt{8})[2(4+n_0U/t)^{-1/4}-(4+n_0U/t)^{1/4}]\psi_{R\uparrow}^\dagger$
and $\alpha_{2,R}=\psi_{R\downarrow}$, with the eigen-energy $\omega_{h2}\!=\!2\sqrt{\smash[b]{4t(4t\!+\!n_0U)}}=2 h_2$
When $h$ deviates slightly from $h_2$, defining $k=k_{R}+q$ and
projecting the original Hamiltonian onto these eigenmodes lead to the effective Hamiltonian:
\begin{align}
    H_R=\omega_{h2}-vq_x\sigma_x-vq_y\sigma_y-\Delta\sigma_z
\label{h2R}
\end{align}
where $ omega_{h2}\!=2 h_2, \Delta=2(h-h_2)$,
$v\!=\!t_\text{s}\sqrt{\smash[b]{2[1\!+\!(8t\!+\!n_0U)/\omega_{h2}\!]}}$. 
Note that the constant term $ omega_{h2} $ shows it is an excited energy which has a direct experimental consequence \cite{CIT}.

Then the dispersion takes the form
\begin{align}
    \omega_{1,2}(q)=\omega_{h2}\mp\sqrt{\Delta^2+v^2(q_x^2+q_y^2)}\>.
\end{align}
Thus the change of band Chern number is $\Delta C_1=-{\rm sgn}(v^2)=-1$.
This analysis is consistent with the numerical result $C_1=0$ at $h>h_2$ and  $C_1=+1$ at $h<h_2$, therefore $0-(+1)=-1$.

At $h=h_1$, one needs to expand the Hamiltonian around $k_X=(\pi,0)$ and $k_Y=(0,\pi)$,
the $T_k$ at $k_X$ and $k_Y$ tells the eigenmodes are
$\alpha_{1,X(Y)}=
(1/\sqrt{8})[2(2+n_0U/t)^{-1/4}+(2+n_0U/t)^{1/4}]\psi_{X(Y)\uparrow}
-(1/\sqrt{8})[2(2+n_0U/t)^{-1/4}-(2+n_0U/t)^{1/4}]\psi_{X(Y)\uparrow}^\dagger$
and $\alpha_{2,X(Y)}=\psi_{X(Y)\downarrow}$,
with the eigen-energy $\omega_{h1}\!=\!2\sqrt{\smash[b]{2t(2t\!+\!n_0U)}}=2h_1+4t$.
When $h$ deviates slightly from $h_1$, defining $k=k_{X(Y)}+q$ and
projecting the original Hamiltonian onto these eigenmodes lead to the effective Hamiltonian
\begin{align}
    H_X&=\omega_{h1}- vq_x\sigma_x+vq_y\sigma_y-\Delta\sigma_z     \nonumber  \\
    H_Y&=\omega_{h1}+ vq_x\sigma_x-vq_y\sigma_y-\Delta\sigma_z
\label{h1XY}
\end{align}
where $ \omega_{h1}\!=2h_1+4t, \Delta=2(h-h_1)$, $v\!=\!t_\text{s}\sqrt{\smash[b]{2[1\!+\!(4t\!+\!n_0U)/\omega_{h1}\!]}}$.
Again, the constant term $ omega_{h2} $ shows it is an excited energy which has a direct experimental consequence \cite{CIT}.

Then the dispersion of $H_X$ or $H_Y$ takes the same form
\begin{align}
    \omega_{1,2}(q)=\omega_{h1}\mp\sqrt{\Delta^2+v^2(q_x^2+q_y^2)}\>.
\end{align}
Thus the change of band Chern number is $\Delta C_1=2{\rm sgn}(v^2)=2$.
This analysis is consistent with the numerical result $C_1=+1$ at $h>h_1$ and  $C_1=-1$ at $h<h_1$, therefore $+1-(-1)=2$.

%Since $\lim\limits_{U\to 0}\!\sqrt{\smash[b]{4t(4t\!+\!n_0U)}}\!=\!4t$
%and $\lim\limits_{U\to 0}\!\sqrt{\smash[b]{2t(2t\!+\!n_0U)}}\!-\!2t\!=\!0$,
Since $\lim_{U\to 0}h_2=4t$ and $\lim_{U\to 0}h_1=0$,
this result suggests that the topology of the Bogoliubov excitation bands at $h>h_c$
is smoothly connecting to that of the non-interacting limit.
Besides, the fact $h_2=\sqrt{4t(4t+n_0U)}>4t$ at any $ U > 0 $
suggests that the region displaying a non-trivial topology ( with a non-zero Chern number) is enlarged with an increasing $U>0$.
%The topological non-trivial region grows as increasing $U>0$ has also been found in the bosonic Haldane model \cite{Ueda2015}.

One can also determine the relation between $h_1$ and $h_c$ when $U/t$ is small.
When $U/t$ is small, $h_1=\sqrt{2t(2t+n_0U)}-2t\sim n_0U/2$, while $h_c\sim n_0U^2t_s^2/t^3 \ll h_1 $.
Thus in the weak coupling limit $ U/t \ll 1 $, there is always a window for $h_c<h<h_{1}$ as shown in Fig.1b.

\begin{figure}[!htb]
    \includegraphics[width=\linewidth]{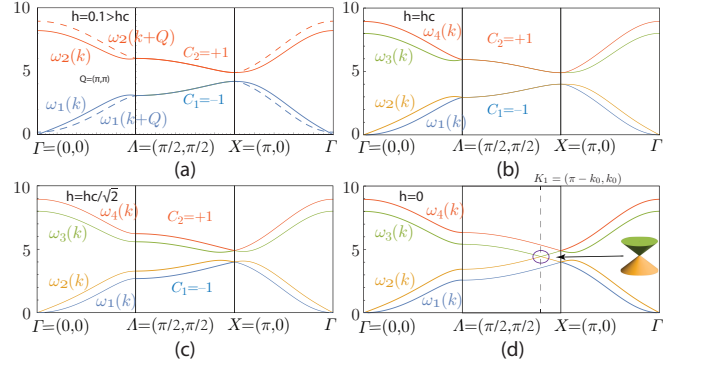}
    \caption{The Non-perturbative  band structure of Bogoliubov excitation bands at %$h\leq h_c$.
    (a) $h>h_c$, (b) $h=h_c$, (c) $h=h_c/\sqrt{2}$, (d) $h=0$.
    The parameters are $t=1$, $t_s=1/2$, $n_0U=1$. It has no analog in the non-interacting limit.
    When $h>0$,  there is always a band gap between two upper bands and two lower bands.
    (a) is just a re-plot of the bottom subfigure of Fig.\ref{Band-h}(b) in the reduced Brillouin zone (BZ),
    so the 2 bands in the BZ changes to 4 bands in the reduced BZ.
    As shown in Sec.C in the SM, two lower (upper) bands also touch along the whole line from $(\pi,0)$ to $(0,-\pi)$
    As shown in (a) and (b), when $h\geq h_c$, they also touch along the whole line from $\Lambda$-point to the X-point along the boundary of the RBZ.
    $ \omega_1 $ is the linear gapless Goldstone mode due to the $ U(1) $ symmetry breaking.
    Note that due to the dropping of the NOFQD effects \cite{NOFQD}, the roton mode $ \omega_1 $ is still a spurious quadratic gapless mode at $ k=0 $.
    Incorporating them back will open a small roton gap at $ k=0 $, %as shown in the solid line in Fig.M4b,
    but will not change the Chern numbers of topology of the band.
    When $h=0$ in (d), there is a Dirac conical band touch at $K_1=(\pi-k_0,k_0)$ of Brillouin zone;
    and the mirror symmetry $ M_1 $  with respect to $k_x=k_y$ tells there is also a Dirac conical band touch at $K_2=(k_0,\pi-k_0)$.
    %(namely still dropping the effects from NOFQD \cite{NOFQD} which transfer the dashed line to the solid line ).
    %Comparing it with (b), they still touch along the whole line from $\Lambda$-point with $ k_\Lambda=(\pi/2,\pi/2) $
    %to the X-point with $ k_X=(\pi,0) $ along the boundary of the reduced BZ,
    %the only crucial difference is that here the roton has a small gap due to the Zeeman field
    %which is the dashed line in Fig.M4b.
    %Note that in contrast to the Z-FM when $ h > h_c $,
    %the $[C_4\times C_4]_\text{D}$ symmetry is broken in the $ N=2 $ XY-AFM phase when $ h > h_c $.
    %(d)This is a re-plot of the bottom subfigure of Fig.\ref{Band-h}(b) in the reduced Brillouin zone, so the 2 bands in the BZ changes to 4 bands in the RBZ.
    %Comparing it with (b), they still touch along the whole line from $\Lambda$-point with $ k_\Lambda=(\pi/2,\pi/2) $
    %to the X-point with $ k_X=(\pi,0) $ along the boundary of the reduced BZ,
    %the only crucial difference is that here the roton has a small gap due to the Zeeman field
    %which is the dashed line in Fig.M4b ( namely still dropping the effects from NOFQD \cite{NOFQD} which transfer the dashed line to the solid line ).
    %It shows that there is no change in the pattern of band Chern numbers across the QPT at $h=h_c$.   }
    }
    \label{Band-0}
\end{figure}

{\sl 6. The topology of the Bogoliubov band when $ 0< h < h_c $ and the absence of the TPT across the QPT at $ h=h_c $.}---
In the $h<h_c$ case,
this corresponds to Eq.(9) and Eq.(10) in Ref.\cite{NOFQD}.
Because of spontaneous translational symmetry breaking,
we have $N=4$ energy bands and $T_k$ is an $8\times8$ matrix.
%We do not know the rigorous way to deal with the nearly order-from-disorder on lattice scale,
%but we expect the corrections from the nearly order-from-disorder do not change the topology.
From the NOFQD analysis in Ref.\cite{NOFQD}, we know the $h=0$ ground-state requires $\theta=\pi/2$,
and $0<h<h_c$ ground-state requires $\theta_h=\arccos(h/h_c)$.
We substitute $\theta=\theta_h$ back into the Eq.\eqref{eq:Hb}, %Eq.M(10) and Eq.M(11),
then calculate the band Chern numbers via Eq.\eqref{eq:C-n}.
In principle, there should be an additional term due to nearly order-from-quantum disorder (NOFQD) correction \cite{NOFQD},
but we expect this correction does not change the topology of the Bogoliubov bands.
From the band structure shown in Fig.\ref{Band-0},
we find that the two lower bands touch, the two upper bands also touch,
and there is always a band gap between the two groups when decreasing $\theta_h$ from $\pi/2$ to $0$
(or equivalently, increasing $h$ from $0$ for $h_c$).
Note that two lower/upper bands touch at the X point $ k_X=( \pi, 0 ) $,
which is a momentum far away from $k=0$,
thus the NOFQD effects ignored so far should not change the band touching behaviors.
Due to the band touching, we need to consider the combination of the Chern numbers $C_{1+2}$ and $C_{3+4}$ when $h<h_c$.
Our numerical evaluation of the integral Eq.\eqref{eq:C-n} gives $(C_{1+2},C_{3+4})=(-1,+1)$.
Recall the result of $h$ just above $h_c$ achieved in the last subsection also gives $(C_1,C_2)=(-1,+1)$,
which is the same pattern as $(C_{1+2},C_{3+4})=(-1,+1)$.
So we conclude that there is not any change in the topology across the quantum phase transition at $ h=h_c $.

{\sl 7. The TPT at $ h =0 $ and the absence of any QPT at $ h=0 $. }---
As shown in Fig.\ref{Band-h},  the topology changes at $ h=h_{1} $ with the conic band crossings at $ k_X=(\pi,0) $ and $ k_Y=(0, \pi) $,
also at $ h=h_{2} $ with the conic band crossing at $ k_R=(\pi, \pi) $. The origin of these TPTs can be traced back and mapped to the
corresponding non-interacting limit of fermions.

In this section, we show that there is also a conic band crossing at $ h=0 $ in Fig.\ref{Band-0}(d) at some
in-commensurate momentum $ (k_{0x}, k_{0y} ) $
between the $\Lambda$-point and X-point, which is not any high symmetry points.
As shown in Fig.\ref{Mag-Dirac}, there are two such Dirac points in the RBZ.
The positions of the two Dirac points are in-commensurate and
also depend on $ t, t_s $ and $ U $, so it comes from the interaction induced OFQD \cite{NOFQD}.
In contrast to the TPTs at $ h= h_1 $ and $ h= h_2 $, the TPT at $ h=0 $  does not have any non-interacting analog.
{\ sl This is also the cental result achieved in this manuscript }.

At $h=0$, the ground state wavefunction is the XY-AFM with $\theta=0$ and $\phi=-\pi/4$.
Solving the band-touching condition $\omega_{2}(k)=\omega_{3}(k)$ leads to the following three cases:
(I) when $n_0U/t<8t_s/(\sqrt{2}t-t_s)$,
two solutions are $K_1=(\pi-k_{0},k_{0})$ and $K_2=(k_{0},\pi-k_{0})$,
%with $k_{0}=\arcsin\Big[\frac{\sqrt{2}t n_0U}{t_s(8t+n_0U)}\Big]<\pi/2$;
with $k_{0}=\arcsin[\sqrt{2}t n_0U/(t_s(8t+n_0U))]<\pi/2$;
(II) when $n_0U/t>8t_s/(\sqrt{2}t-t_s)$,
two solutions are $P_{1,2}=\pm(p_{0},p_{0})$,
%with $p_{0}=\arcsin\Big[\frac{4t}{8t+n_0U}\sqrt{\frac{4t(4t+n_0U)}{4t^2-2t_s^2}}\Big]<\pi/2$;
with $p_{0}=\arcsin[(4t/(8t+n_0U))\sqrt{4t(4t+n_0U)/(4t^2-2t_s^2)}]<\pi/2$;
(III) when $n_0U/t=8t_s/(\sqrt{2}t-t_s)$,
the two solutions merge into just one at  $\Lambda=(\pi/2,\pi/2)$.
Intuitively, as shown in Fig.S2, when keeping $h=0$,
increasing $n_0U/t$ from $0^+$ to $8t_s/(\sqrt{2}t-t_s)$,
the $\omega_{2,3}$ conically touch at $K_{1,2}$;
then at $n_0U/t=8t_s/(\sqrt{2}t-t_s)$, $K_1$ and $K_2$ collide at $\Lambda$-point;
further increasing $n_0U/t$, they bounce off along the two opposite directions along the perpendicular direction \cite{footnote},
so $\omega_{2,3}$ conic touches at $P_{1,2}$.

Around $h=0$, defining $k=K_{1,2}+q$ or $k=P_{1,2}+q$,
we can similarly expand the Hamiltonian around the two minima of $\omega_{2,3} $ to
find the effective Hamiltonian:
\begin{align}
%    H_{1}=\omega_{1q}\!+v_1(q_x-q_y)\sigma_x+v_2(q_x+q_y)\sigma_y\!-\Delta\sigma_z\>, \nonumber\\
%    H_{2}=\omega_{2q}\!-v_1(q_x-q_y)\sigma_x-v_2(q_x+q_y)\sigma_y\!-\Delta\sigma_z\>,
    H_{1}=\omega_{0}+c_\parallel q_\parallel
            +v_\parallel q_\parallel\sigma_x
            +v_\perp q_\perp\sigma_y
            \!-\Delta\sigma_z\>, \nonumber\\
    H_{2}=\omega_{0}-c_\parallel q_\parallel
            -v_\parallel q_\parallel\sigma_x
            -v_\perp q_\perp\sigma_y
            \!-\Delta\sigma_z\>,
\label{H12}
\end{align}
where $\omega_{0}=8t(4t+n_0U)/(8t+n_0U)$,
$q_\parallel=(q_x-q_y)/\sqrt{2},q_\perp=(q_x+q_y)/\sqrt{2}$,
and $\Delta=hn_0U\sqrt{t/(4t+n_0U)}$ for regime (I),
$q_\parallel=(q_x+q_y)/\sqrt{2},q_\perp=(q_x-q_y)/\sqrt{2}$,
and $\Delta=4htt_s/\sqrt{2t^2-t_s^2}$ for regime (II).
%gap parameters are $\Delta=hn_0U\sqrt{t/(4t+n_0U)}$ for regime (I),
%and $\Delta=4htt_s/\sqrt{2t^2-t_s^2}$ for regime (II).
The effective velocities $v_\parallel,v_\perp,c_\parallel>0$ are proportional to $\cos(k_0)$ or $\cos(p_0)$ for regime (I) and regime (II) respectively
and are listed in the appendix B.

 The dispersion of $H_i, i=1,2$ takes the form
\begin{align}
    \omega_{2,3}(q)=\omega_0-(-1)^i c_\parallel q_\parallel
    \mp\sqrt{\Delta^2+v_\parallel^2q_\parallel^2+v_\perp^2q_\perp^2}\>.
\label{K12}
\end{align}
It is necessary to stress that there exists a Doppler shift term, the $c$ term, in $H_{1,2}$,
which is the salient feature unique to the OFQD induced TPT, not shared by any non-interacting counter-part.
Of course, it does not affect the value of band Chern numbers.
%because the two minima are not located at any high symmetry points,

%the band Chern number of $\omega_1$ and $\omega_2$ is $C_{1+2}=-{\rm sgn}(v_1v_2\Delta)$,
%and
From the effective Hamiltonian Eq.\eqref{H12},
Eq.\eqref{C-} tells the change of band Chern number is $\Delta C=-2 {\rm sgn}(v_\parallel v_\perp)=-2$
when $\Delta$ changes from positive to negative.
This result is consistent with the numerical evaluation via Eq.\eqref{eq:C-n},
which gives $C_{1+2}=-1$ at $h>0$ and $C_{1+2}=+1$ at $h<0$, therefore $ -1-(+1)=-2 $.

At $ h=0 $, in the colliding point $n_0U/t=8t_s/(\sqrt{2}t-t_s)$,
the $K_1$ and $K_2$ merge at $\Lambda=(\pi/2,\pi/2 )$-point.
Expanding around the $\Lambda$-point leads to the dispersion:
\begin{align}
 \omega_{2,3}(q)=\omega_0(q) \mp \sqrt{u_1^2 q^4_\perp +u_2 q^4_{\parallel} +w q^2_{\parallel} q^2_\perp },
\label{Klambda}
\end{align}
%$\omega_{2,3}(q)=\omega_0(q) \mp \sqrt{u_1^2 q_-^4+u_2^2 q_+^4+w q_-^2 q_+^2}$,
where $u_1=t^2/(\sqrt{2}t_s)$,
$u_2=t^2t_s/[\sqrt{2}(2t^2-t_s^2)]$,
$w=t^2(t^2-t_s^2)/(2t^2-t_s^2)$,
and $ \omega_0(q)=4t+2\sqrt{2}t_s + DS $ where one can see the Doppler shift term
$ DS= [t_s/\sqrt{2}][-q^2_{\parallel}+ q^2_\perp t_s^2/(2t^2-t_s^2)] $  also gets to the quadratic order.

Note that $4u_1^2u_2^2>w^2$ is ensured by the condition $0<t_s<\sqrt{2}t$,
so the square root is always positive define.
Since $w$-term won't lead to a gap close, one can ignore $w$-term and rescale $q$
to make $\omega_{2,3}(q)$ isotropic.
The $\omega_{2,3}(q)\sim q^2$ is consistent with the fact that $v_\parallel$, $v_\perp$, $c_\parallel$ in Eq.\eqref{H12}
vanish as $k_0$ or $p_0$ approaching $\pi/2$.
When $h$ deviates from $0$, the effective Hamiltonian belongs to $n=2$ case of Eq.\eqref{Hn}.
Thus, although there is only one band touching point with the monopole charge $ n=2 $,
the change of Chern number is still $-2$.
%This result is also consistent with numerical evaluation of Eq.\eqref{eq:C-n},
%which gives $C_{1+2}=-1$ at $h>0$ and then $C_{1+2}=+1$ at $h<0$.

In short, regardless of the value of $n_0U$,
the change of the Chern number from $h=0^+$ to $h=0^-$ is always $-2$.
So the scattering process in Fig. S1, is not a TPT.

{\sl 8. Dramatic differences between the non-interacting fermionic band structure and
             the bosonic Bogoliubov band structure at the quadratic level.}---
   It is important to stress some crucial differences between the
   fermionic band structure and the bosonic Bogliubov band structure at the quadratic level and also pointed out the common mistakes
   made in all the previous literatures to calculate the  edge modes associated with the bosonic Bogliubov band structure:

   In the fermionic or bosonic QAH problem,  the Time reversal symmetry is broken,
   the Chern number is not protected by any symmetry, its definition involves no symmetry requirement.
   The former is really non-interacting, while the bosonic Bogoliubov band is non-interacting at the quadratic level only
   where one dropped the cubic and quartic interactions and all the higher order interactions.
   All these interactions are not important except near the QPT at $ h=h_c $.
   In fact, as shown by the RG analysis, near the QPT, the two quartic interaction in Eq.M32 are marginally irrelevant.
   But both are relevant inside the XY-AFM phase and in fact, leads to the symmetry breaking inside the phase.

   For the fermionic problem, there are always edge states associated with the  topological phases.
   This is justified, because a single fermion operator never condenses.
   One tempts to also calculate the edge modes
   within the band gaps in Fig.\ref{Band-h} and Fig.\ref{Band-0} associated with the bosonic Bogoliubov band.
   Unfortunately, this kind of calculation has mathematical meanings,
   but  makes no physical sense: this is because Eq.M3 is based on the BEC at $ k=0 $  in the Z-FM phase,
   Eq.M9 with the parametrization Eq. M11 are based on the BEC at both $ k=0 $ and $ Q=(\pi,\pi) $
   in the XY-AFM phase. Unfortunately, they are ill-defined in a strip geometry.
   Ref.\cite{Ueda2015} and many other previous works,  just copied the same method
   from its fermionic counterpart to evaluate edge modes associated with the bosonic Bogoliubov band:
   They simply transfer the  quadratic Bogoliubov band in
   momentum space to real space, then solve the edge modes in the strip geometry by putting a periodic boundary condition
   along one direction, then the open boundary condition along the other.
   Unfortunately, this approach is well-planned mathematically, but is not self-consistent physically, because it ignored the root of BEC which
   leads to the quadratic bands in the first place. As stressed above, the BEC root
   is ill-defined in the strip geometry. So it makes no physical sense to talk about the edge modes.
   The claims made in some experimental works to measure the edge modes have no physical grounds.

{\sl 9. Experimental detections.}---
  So far, the experiment\cite{2dsocbec} has detected the non-interacting fermionic Chern numbers Fig.1a
  using the highly excited bosonic spinors, so it just mimic the single particle properties of the Hamiltonian Eq.1
  using a spinor boson with SOC.
  Its main purpose is to demonstrate a possible realization of the bosonic Hamiltonian Eq.1 using cold atoms.
  The common drawbacks of most cold atom experiments is just to demonstrate a possibility to simulate  a Hamiltonian
  which has been well studied in materials and claim the ability to tune the parameters.
  But in reality, it is rare to go beyond just a simulation  to  demonstrate new many-body phenomenon or topological phenomena due to many-body interactions..

  Here we showed that there are two kinds of  topological band structures
  (1) The one in the regime $ | h| > |h_1 | $ is smoothly connected to the non-interacting fermionic one.
  So it can also be called perturbative regime.
   Even so, the two critical fields $ h_{1,2} $ listed in Eq.\ref{h12} still depends on the interaction $ n_0 U $.
   All the parameters in the effective Hamiltonian Eq.\ref{h2R} and \ref{h1XY} also depend on the interaction $ n_0 U $.
  (2) The one in the  regime $ -h_1 < h < h_1 $ has no analog in the non-interacting counter-part, so it is
      completely due to the many-body interaction. More specifically, due to the non-perturbative OFQD.
      So it can also be called non-perturbative regime. It is also an experimentally easily accessible regime in the cold atom systems.

  It is extremely important to push the already 6 year old experiment\cite{2dsocbec} beyond the single-particle picture:
  to detect this novel purely interaction induced topological phenomena.
  In contrast to the detections suggested in \cite{NOFQD} which are all equilibrium properties in the ground states and low energy excitations, 
  here are the excited states near the high  energy $ \omega_{h2},   \omega_{h1} $ and $ \omega_0 $ in Eq.\ref{h2R},\ref{h1XY} and \ref{H12} 
  near the high momentum $ R=(\pi,\pi) $ or $ X= (\pi,0), Y=(0,\pi) $ and tunable in-commensurate momenta $ K_{1,2} $ or $ P_{1,2} $ respectively \cite{CIT}.
  We suggest to also measure the scattering process of the two bosonic Dirac points at $ h=0 $ displayed in Fig.S1 by the 
  Bragg spectroscopy \cite{becbragg,bragg1,bragg2}  or the momentum resolved interband transitions \cite{mrit} adapted to the excited states.

{\sl 10. Conclusions.}---
  For any interacting Hamiltonian with a non-trivial topology in the non-interacting limit such as the QAH one Eq.1,
  there are always two aspects quantum and topology.
  The former focus on the ground states and quantum phase transitions which may only depend on the low energy excitations
  around the band minima in the weak interaction limit. OFQD is needed to even determine the ground state.
  NOFQD is needed to determine the QPT at $ h=h_c $.
  The latter focus on the global structure of the bands, therefore also  the excited states to capture the global topology \cite{CIT}.
  However, the Bogliubov quasi-particle band picture breaks down near the QPT near $ h=h_c $ where a RG analysis is needed to
  capture all the physics well beyond the quadratic band picture.
  
 However, as classified in \cite{NOFQD}, there are two kinds of OFQD, the first response trivially to
 a deformation, no NOFQD phenomenon emerging from the OFQD. The second response highly non-trivially to
 a deformation and lead to  NOFQD phenomenon. It would be interesting to see if there are also two classes here, the first leads to a TPT, as is the case
 presented in this work, the second does not lead to any TPT. The two criterions coincide in the present case.
 It is also worthy to see if the two criterions coincide in other cases.

 The OFQD and the topological invariants such as Chern number seem are two very different concepts.
 The first concept is a completely many-body quantum phenomenon. While the latter is mainly a non-interacting topological phenomenon.
 Here we show that
 there are deep connections between the two in the context of the experimentally realized weakly interacting Quantum Anomalous Hall system
 of spinor bosons in an optical lattice. We expect this new effects could also be realized in
 frustrated quantum spin systems which leads to topological phase transitions of magnons.

{\bf Acknowledgements }

 J. Ye thanks Prof. Gang Tian for the hospitality during his  visit at  the School of Science of Great Bay University.

%We thank B. Halperin for a critical comment \cite{bert} which inspired us to discover the NOFQD.

%We thank Prof. Congjun Wu and Prof. Gang Tian for the hospitality during our visit at
%the Institute for Theoretical Sciences, Westlake University, Hangzhou, 310024, Zhejiang, China.

\onecolumngrid
%%%%%%%%%% Merge with supplemental materials %%%%%%%%%%
%\pagebreak
\widetext
\begin{center}
\textbf{\large
Supplemental Materials:
Topological phase transitions induced by order from quantum disorder}
\end{center}
%%%%%%%%%% Merge with supplemental materials %%%%%%%%%%
%%%%%%%%%% Prefix a "S" to all equations, figures, tables and reset the counter %%%%%%%%%%
\setcounter{equation}{0}
\setcounter{figure}{0}
\setcounter{table}{0}
\setcounter{page}{1}
\makeatletter
\renewcommand{\theequation}{S\arabic{equation}}
\renewcommand{\thefigure}{S\arabic{figure}}
\renewcommand{\bibnumfmt}[1]{[S#1]}
\renewcommand{\citenumfont}[1]{S#1}
%%%%%%%%%% Prefix a "S" to all equations, figures, tables and reset the counter %%%%%%%%%%

 In this Supplemental Materials we provide some details of the band structures which are needed to construct the
 effective Hamiltonian near the various TPTs.

\section{ A. Symmetry analysis and the scattering process of the two bosonic  Dirac points at $ h=0 $ }

%We are following the convention given by Dresselhaus's Book,
The high symmetry points in the 1st Brillouin zone are defined as
\begin{align}
    &\Gamma=(0,0),~\Lambda=(\pi/2,\pi/2),~%\nonumber\\
    R=(\pi,\pi),~ X=(\pi,0),~ Y=(0,\pi).
\end{align}
Two additional special $k$-points are defined as the band minima ( maxima ) of $\omega_3(k)$  ( $\omega_2(k)$ ) when $|h|<h_c$,
they are functions of spin-orbit coupling $ t_s $ and the interaction $ U $.
%When choosing $c_1=\cos(\theta_h/2),c_2=e^{-i\frac{\pi}{4}}\sin(\theta_h/2)$,

At $h=0$,  if the interaction strength $ U $ is sufficiently small $n_0U/t<8t_s/(\sqrt{2}t-t_s)$,
the $\omega_3(k)$ has two minima at
$k=K_1=(\pi-k_{0},k_{0})$ and $k=K_2=(k_{0},\pi-k_{0})$,
where
\begin{align}
    k_{0}=\arcsin\Big[\frac{\sqrt{2}t n_0U}{t_s(8t+n_0U)}\Big]\>.
\end{align}
%When increasing $h$ form $0$ to $h_c$, the $k_{0}$ will gradually become $0$,
%thus $K_1$ becomes $Y$-point and $K_2$ becomes $X$-point.
When $n_0U$ goes to $0$, the $k_{0}$ will become $0$,
thus $K_1$ becomes $X$-point and $K_2$ becomes $Y$-point.

At $h=0$, if the interaction strength $ U $ is relatively large $n_0U/t>8t_s/(\sqrt{2}t-t_s)$,
the $\omega_3(k)$ has two minima at
$k=P_1=(+p_{0},+p_{0})$ and $k=P_2=(-p_{0},-p_{0})$,
where
\begin{align}
    p_0=\arcsin\Big[\frac{4t}{8t+n_0U}\sqrt{\frac{4t(4t+n_0U)}{4t^2-2t_s^2}}\Big]
\end{align}

At $h=0$, if  $n_0U/t=8t_s/(\sqrt{2}t-t_s)$,
the $\omega_3(k)$ has only one minima at $k=\Lambda=(\pi/2,\pi/2)$ with the monopole charge $ n=2 $.

\begin{figure}[!htb]
    \includegraphics[width=0.29\linewidth]{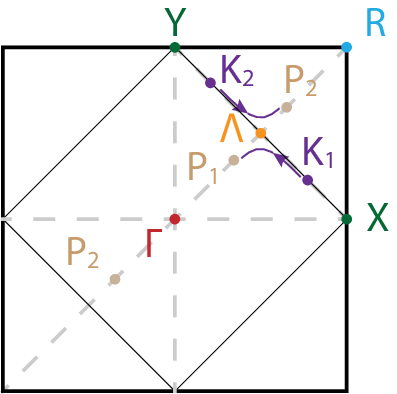}
    \caption{Illustration of the special k-points used in this Letter:
    $\Gamma=(0,0),~\Lambda=(\pi/2,\pi/2),~R=(\pi,\pi),~ X=(\pi,0),~ Y=(0,\pi)$.
    The $K_1$ and $K_2$ are the two minima of $\omega_3(k)$ when $n_0U/t<8t_s/(\sqrt{2}t-t_s)$;
    the $P_1$ and $P_2$ are the two minima of $\omega_3(k)$ when $n_0U/t>8t_s/(\sqrt{2}t-t_s)$.
    The arrows show that: as  $n_0U/t$ increases,
    the two minima of $\omega_3(k)$  approach each other along the X-Y line, then collide at $\Lambda$-point,
    then bounce off into two opposite  perpendicular directions.
    The thick square is the lattice Brillouin Zone (BZ), the thin one is the reduce Brillouin Zone (RBZ).   }
    \label{KP}
\end{figure}

\begin{figure}[!htb]
    \includegraphics[width=0.55\linewidth]{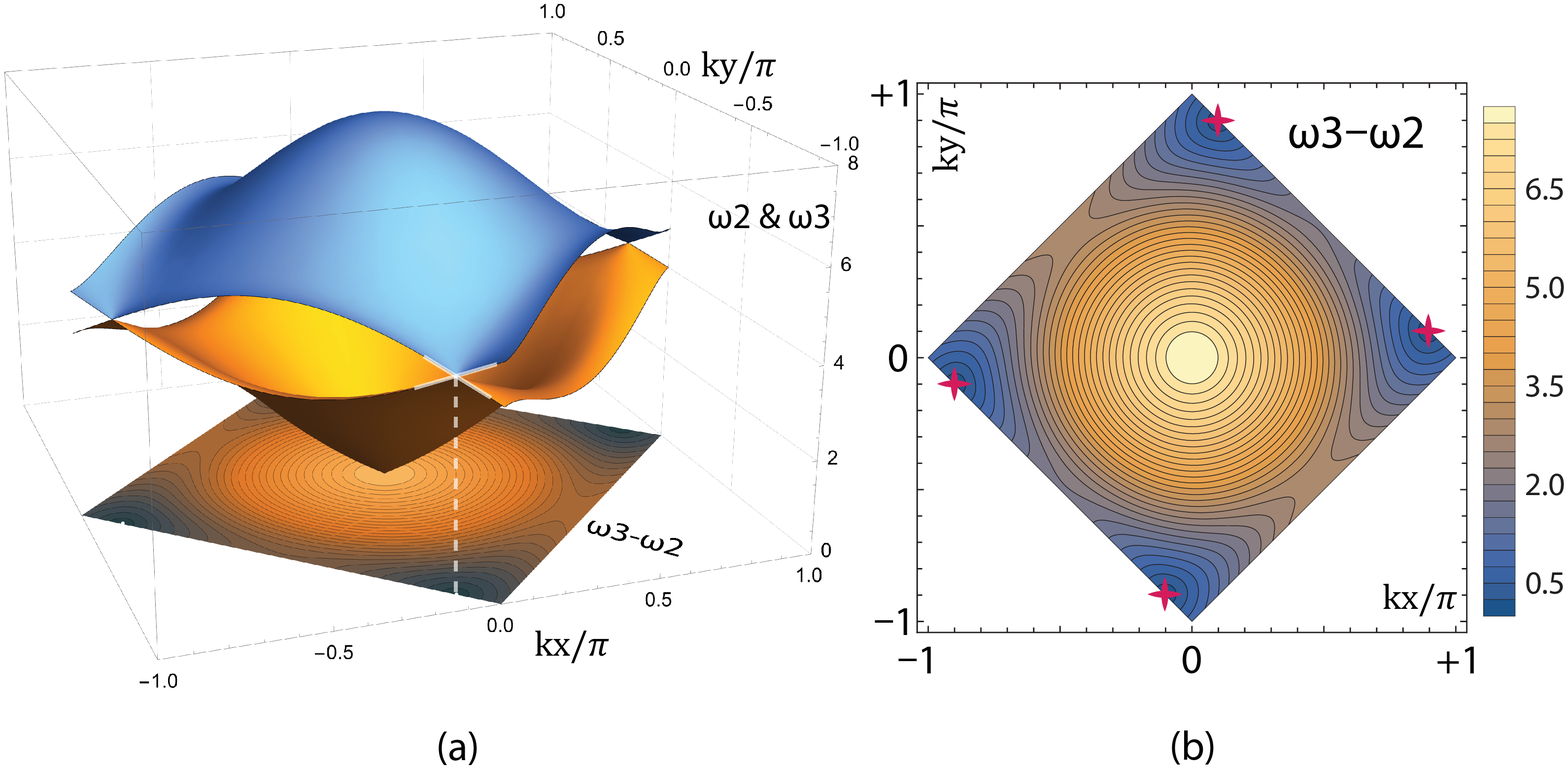}
    \caption{The OFQD induced band structure of Bogoliubov excitation bands $\omega_2$ and $\omega_3$ at $h=0$.
    The parameters are $t=1$, $t_s=1/2$, $n_0U=1$, and %$(c_0,c_\pi)=(1,e^{-i\pi/4})/\sqrt{2}$.
    $\theta=\pi/2$, $\phi=-\pi/4$.
    The bottom plane is the contour plot of $\omega_3-\omega_2$ as a function of $k_x$ and $k_y$.
    It is clear to see there are two Dirac points at the two in-commensurate momenta denoted by $+$  sitting on the Reduced BZ boundary.  }
    \label{Mag-Dirac}
\end{figure}

In summary, pictorially as illustrated in Fig.\ref{KP}, at $h=0$,
increasing $n_0U$ from $0^+$ to the value greater than $8tt_s/(\sqrt{2}t-t_s)$,
the two minima of $\omega_3(k)$ approach each other along the line $k_{0x}+k_{0y}=\pi$,
then collide at $\Lambda$-point, then the two minima bounce off along the opposite direction along the perpendicular direction
$k_{0x}=k_{0y}$.

The point group symmetry of the square lattice is $C_{4v}$.
In the presence of spin-orbit coupling,
the $C_{4v} $ symmetry becomes $[C_4\times C_4]_\text{diag}$ symmetry,
which is a joint 4-fold rotation of spin and orbit.
Besides, there are also two mirror symmetries $M_{1,2}$,
which also inherit from the $C_{4v}$ point group symmetry.
The mirror symmetry $M_1$
is a composition of time reversal $ i \to -i, \vec{\sigma} \to -\vec{\sigma} $, the orbit reflection $k_x\leftrightarrow k_y$,
and the spin rotation $(\sigma_x,\sigma_y,\sigma_z)\to(\sigma_y,\sigma_x,-\sigma_z)$.
The mirror symmetry $M_2$
is a composition of the orbit reflection $ \vec{k} \to  -\vec{k} $,
and the spin rotation $(\sigma_x,\sigma_y,\sigma_z)\to(-\sigma_x,-\sigma_y,\sigma_z)$.

When $ h > h_c $, the ground state is the Z-FM superfluid phase where both $[C_4\times C_4]_\text{diag}$ and $ M_{1,2} $ symmetries are unbroken.

When $ h < h_c $, it is the XY-CAFM superfluid where the $[C_4\times C_4]_\text{diag} \to 1 $ is broken while $M_{1,2} $ remain unbroken.
It is the $M_1$ symmetry which guarantees the two minima $K_1=(k_{0},\pi-k_{0})$ and $K_2=(\pi-k_{0},k_{0})$
are related by $k_x\leftrightarrow k_y$.
It is the $M_2$ symmetry which guarantees the two minima $P_1=(+p_{0},+p_{0})$ and $P_2=(-p_{0},-p_{0})$
are related by $k\leftrightarrow -k$. The colliding point  $ \Lambda=(\pi/2, \pi/2 ) $ is the mirror symmetric point which
is invariant under both $ M_1 $ and $ M_2 $.  

It is constructive to compare the mirror symmetric point  $ \Lambda=(\pi/2, \pi/2 ) $   in the momentum space here with that
in the SOC parameter space $ \beta=\pi/2 $ discussed in \cite{SOCM}. 
In the former, the Hamiltonian which is the sum over all momenta  has the same symmetry, the Doppler shit term
still exists at a quadratic order, but non-vanishing.
While in the latter, the Hamiltonian has an enlarged symmetry which dictates the exact vanishing of the Doppler shift term.

It is interesting to compare the scattering process of the two bosonic Dirac points with the same chirality  in Fig.\ref{KP} with
the annihilation process of the two fermionic Dirac points with the opposite chirality discussed in \cite{TQPT}.
As shown in this work, the former happens in the high energy modes, so it is not a QPT. It is not a TPT either.
While the latter is both where various physical quantities satisfy various scaling functions.
The collision of two bosonic vortices with opposite vorticities in an expanding curved space-time was analyzed in \cite{gravity} where
two surface waves also emerge after the collision and head in opposite direction along the normal direction.

% \clearpage
\section{ B. Determination of the band structures  induced by the OFQD at $h=0$. }

%\subsection{band structures at $h=0$}
Solving the band touching condition $\omega_{2}(k)=\omega_{3}(k)$
within the regime (I) $n_0U/t<8t_s/(\sqrt{2}t-t_s)$, %$c_0=1/\sqrt{2}$ and $c_\pi=e^{-i\pi/4}/\sqrt{2}$,
%and also assuming the interaction strength is sufficiently small,
we obtain the two solutions $K_1=(k_{0x},k_{0y})$ and $K_2=(k_{0y},k_{0x})$,
where
\begin{align}
    k_{0y}=\arcsin\Big[\frac{\sqrt{2}t n_0U}{t_s(8t+n_0U)}\Big]=k_0,\quad
    k_{0x}=\pi-k_{0}.
\end{align}
The two solutions are related by the remaining $M_1$ symmetry.
The band touching happened at high energy
$\omega_{2}(K_1)=\omega_{3}(K_1)=8t(4t+n_0U)/(8t+n_0U)=\omega_{0}> 4t$.
Expanding the $\omega_{2,3}(k)$ near $K_{1,2}$ with $k=K_{1,2}+q$, we obtain
\begin{align}
    \omega_{2,3}(K_{1}+q)
        =\omega_{0}\mp\sqrt{v_1^2 (q_x-q_y)^2+v_2^2 (q_x+q_y)^2}+c(q_x-q_y)\>,  \nonumber  \\
    \omega_{2,3}(K_{2}+q)
        =\omega_{0}\mp\sqrt{v_1^2 (q_x-q_y)^2+v_2^2 (q_x+q_y)^2}-c(q_x-q_y)\>,
\end{align}
where
\begin{align}
  v_1 & =\sqrt{2}\cos k_0 \, t_s(8t+n_0U)/\sqrt{\smash[b]{16t(4t+n_0U)}},   \nonumber  \\
  v_2 & =\sqrt{2}\cos k_0 \, t_s(8t+n_0U)^2/[16t(4t+n_0U)],    \nonumber  \\
   c  & =\sqrt{2}\cos k_0\, t_s n_0^2U^2/[16t(4t+n_0U)].
\label{threevelocitiesK}
\end{align}
The two solutions are related by the remaining $M_1$ symmetry.
The Doppler shift term is along the $[1\bar{1}]$ direction,
which is nothing but the ``collision'' direction in Fig.\ref{KP}.
Note that the relative sizes of $ v_1 $ and $ c $ is not important here, due to the high energy $ \omega_0 $ which has direct experimental consequences
as stressed in Sec.M9, so
it will not drive any instabilities. In a sharp contrast, in the fermionic QAH under an injecting current discussed in \cite{QAHinj},
due to the absence of the  $ \omega_0 $ term, the sign change of the relative size drives a TPT.

Solving the band touching condition $\omega_{2}(k)=\omega_{3}(k)$
within the regime (II) $n_0U/t>8t_s/(\sqrt{2}t-t_s)$,
we obtain the two solutions $P_1=+(k_{0x},k_{0y})$ and $P_2=-(k_{0x},k_{0y})$,
where
\begin{align}
    k_{0x}=k_{0y}=\arcsin\Big[\frac{4t}{8t+n_0U}\sqrt{{\frac{4t(4t+n_0U)}{4t^2-2t_s^2}}}\Big]=p_0.
\end{align}
The two solutions are related by the remaining $M_2$ symmetry.
The band touching happens at the high energy
$\omega_{2}(P_1)=\omega_{3}(P_1)=8t(4t+n_0U)/(8t+n_0U)=\omega_{0}> 4t+2\sqrt{2}t_s$.
Expanding the $\omega_{2,3}(k)$ near $P_{1,2}$ with $k=P_{1,2}+q$, we obtain
\begin{align}
    \omega_{2,3}(P_{1}+q)
        =\omega_{0}\mp\sqrt{v_1^2 (q_x+q_y)^2+v_2^2 (q_x-q_y)^2}+c(q_x+q_y)\>,  \nonumber  \\
    \omega_{2,3}(P_{2}+q)
        =\omega_{0}\mp\sqrt{v_1^2 (q_x+q_y)^2+v_2^2 (q_x-q_y)^2}-c(q_x+q_y)\>,
\end{align}
  where
\begin{align}
  v_1 & =[2t(8t+n_0U)/(n_0U)]\cot p_0,   \nonumber  \\
  v_2 & =[t_s(8t+n_0U)/\sqrt{8t(4t+n_0U)}]\cos p_0,   \nonumber  \\
   c  & =[2t n_0U/(8t+n_0U)]\cot p_0
\label{threevelocitiesP}
\end{align}
The Doppler shift term is along the $[11]$ direction,
which is nothing but the ``bouncing-off'' direction in Fig.\ref{KP}.

In summary, near the band touching point, we always have
\begin{align}
    \omega_{2,3}(K_{i}+q \text{ or } P_{i}+q)
        =\omega_{0}\mp\sqrt{\smash[b]{v_\parallel^2 (q_\parallel)^2+v_\perp^2 (q_\perp)^2}}-(-1)^i c_\parallel q_\parallel\>,
\end{align}
where $i=1,2$, $q_\parallel=(q_x-q_y)/\sqrt{2},q_\perp=(q_x+q_y)/\sqrt{2}$ for regime (I),
$q_\parallel=(q_x+q_y)/\sqrt{2},q_\perp=(q_x-q_y)/\sqrt{2}$ for regime (II),
and $v_\parallel,v_\perp,c_\parallel$ are easy to deduce from the previous expressions
Eq.\ref{threevelocitiesK} and Eq.\ref{threevelocitiesP}.

%\clearpage

\section{ C. The folding of the two lower bands and the two upper bands and
the physical meanings of $ C_{I}= C_{1+2}$ and $ C_{I\!I}=C_{3+4}$ at $ h < h_c $. }

\begin{figure}[!htb]
    \includegraphics[width=0.55\linewidth]{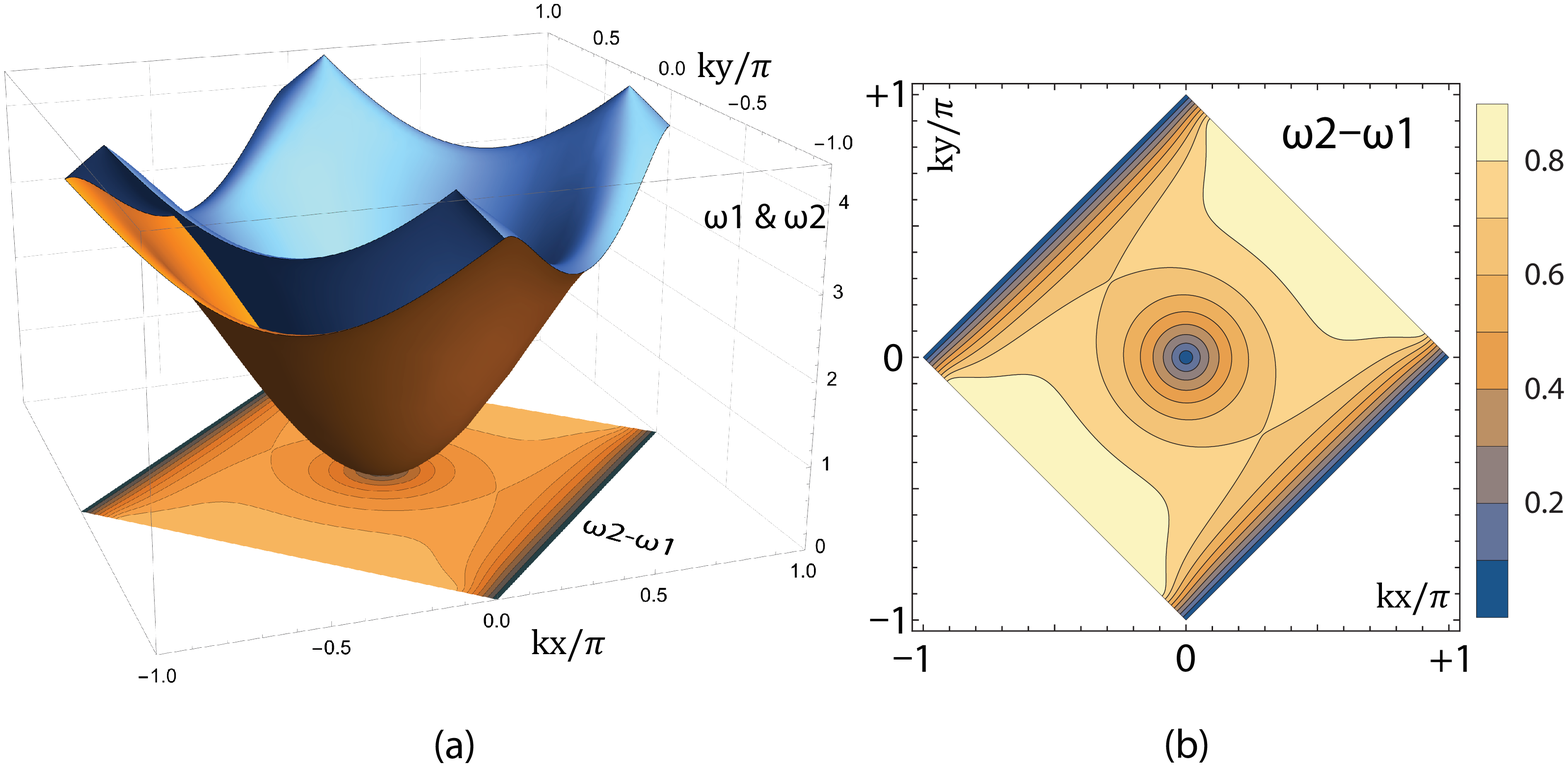}
    \caption{The band structure of the Bogoliubov excitation bands $\omega_1$ and $\omega_2$ at $h=0$.
    The bottom plane is the contour plot of $\omega_2-\omega_1$ as a function of $k_x$ and $k_y$.
    It is clear to see the two bands merge along one reduced BZ edge from $ (0, -\pi) $ to $ (\pi,0) $.
    The parameters are the same as Fig.\ref{Mag-Dirac}.   }
    \label{Band12}
\end{figure}

 Now we look at the band structures of the two lower bands and the two upper bounds.
At $ h=0 $, in the XY-AFM phase with $c_0=1/\sqrt{2}$ and $c_\pi=e^{-i\pi/4}/\sqrt{2}$,
solving the band touching condition $\omega_{1}(k)=\omega_{2}(k)$ or $\omega_{3}(k)=\omega_{4}(k)$,
we find the solutions $k=(k_{0x},k_{0y})$ form a straight line $\{(k_{0x},k_{0y})|k_{0x}-k_{0y}=\pi\}$.
Thus, there is a line degeneracy between $\omega_{1}$ and $\omega_{2}$.
The Fig.\ref{Band12} shows that the $\omega_{1}$ and $\omega_{2}$ indeed touch along one reduced BZ edge.

The line degeneracy between $\omega_{1,2}$ and $\omega_{3,4}$ suggests that
we can unfold the two bands along $[1\,\bar{1}]$ direction,
by defining $\Omega_{I}(k)=\omega_2(k)$ if $k$ in reduced BZ,
otherwise $\Omega_{I}(k)=\omega_1(k+Q')$ where $Q'=(+\pi,-\pi)$.
Similarly, we define $\Omega_{I\!I}(k)=\omega_3(k)$ if $k$ in the reduced BZ,
otherwise $\Omega_{I\!I}(k)=\omega_4(k+Q')$.

\begin{figure}[!htb]
    \includegraphics[width=0.55\linewidth]{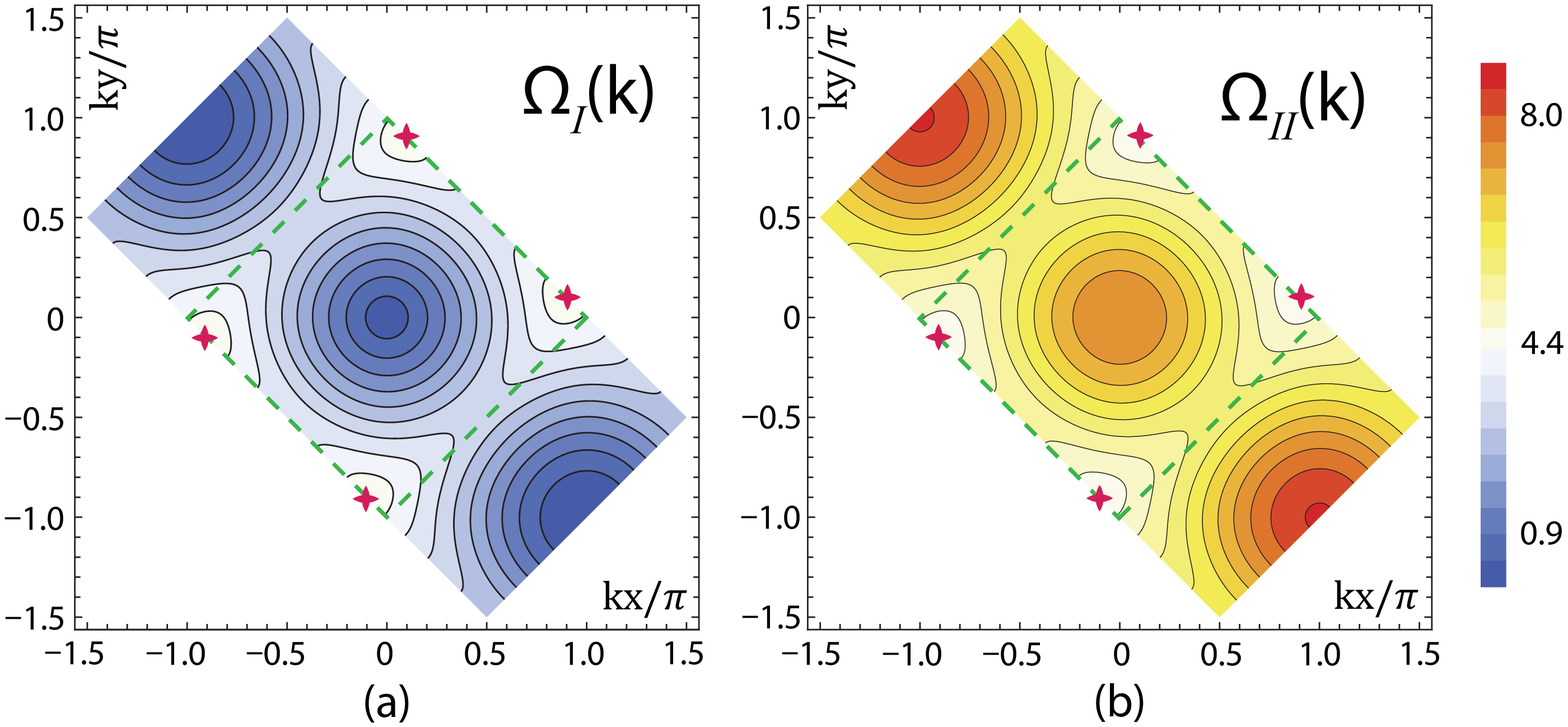}
    \caption{The unfold Bogoliubov excitation bands $\Omega_{I}$ and $\Omega_{I\!I}$ at $h=0$.
    There are two conic band crossings between $\Omega_{I}$ and $\Omega_{I\!I}$.
    The green dashed line represents the reduce Brillouin zone.
    The new Brillouin zone is a rectangular shape, and the two Dirac points denoted by $+$ sit on the longer side.
    The parameters are the same as Fig.\ref{Mag-Dirac}.
    When adding a small $0<|h|<h_c$, the line degeneracy still exists,
    so $\Omega_{I}(k)$ and $\Omega_{I\!I}(k)$ are still well-defined. Due to the symmetry $ [C_4 \times C_4 ]_D $  restoration of the Z-FM above $ h>h_c $,
    the line degeneracy remains.
    Of course, any  small $h$ opens a gap between $\Omega_{I}$ and $\Omega_{I\!I}$ around the two Dirac points and Lead to
    the two effective Hamiltonians Eq.M13.   }
    \label{Omega}
\end{figure}

The unfolded band structures are plotted in Fig.\ref{Omega}.
Thus it is more nature to consider $\omega_{1,2}$ as one band and $\omega_{3,4}$ as the other band;
this is  the physical  reason we always consider a combination of band Chern numbers
$ C_{I}= C_{1+2}$ and $ C_{I\!I}=C_{3+4}$.

\begin{thebibliography}{99}
\bibitem{footnote} Our calculations are only valid in the regime $t_s/t<\sqrt{2}$ and $n_0U/t$ below the SF-MI transition.
If we fix $t_s$ to be some typical value between 0 and $\sqrt{2}t$,
i.e. $t=1$ and $t_s=1/2$, then $8t_s/(\sqrt{2}t-t_s)\approx 4.38$,
while the typical $(U/t)_c$ in the Bose Hubbard model at $n=1$ for the SF-MI  is about $16$.
Thus both (I) and (II) are reachable in teh SF phase.
\end{thebibliography}
\end{document}